\newcommand{\be}{\begin{eqnarray}}
\newcommand{\ee}{\end{eqnarray}}
\def\l({\left(}
\def\r){\right)}
\def\rr{{\bb R}}
\def\vs#1{\vskip#1}
\def\ni{\noindent}

\def\bd{\begin{displaymath}}
\def\ed{\end{displaymath}}
\def\ba#1{\begin{array}{#1}}
\def\ea{\end{array}}
\def\nn{\nonumber}
\newfont{\Bbb}{msbm10 scaled 1200}
\newfont{\bb}{msbm10 scaled 1200}
\newfont{\Bbbs}{msbm10 scaled 1000}
\newfont{\Bbbl}{msbm10 scaled 1440}
\documentstyle[12pt,fleqn]{article}
\title{Calogero model and $sL(2$,{\Bbbl R}) algebra}
\author{Cezary Gonera\thanks{supported by KBN grant 2 P03B 076 10}
\ \ \ Piotr Kosi\'nski$^*$\\
Department of Theoretical Physics II\\
University of \L\'od\'z\\
Pomorska 149/153\\
90--236 \L\'od\'z, POLAND}
\date{}
\begin{document}
\maketitle
\begin{abstract}
The Calogero model with external harmonic oscillator potential is
discussed from 
$sL(2$,{\Bbbs R}) algebra point of view. Explicit formulae for
functions with 
exponential time behaviour are given; in particular, the integrals of
motion 
are constructed and their involutivness demonstrated. The
superintegrability 
of the model appears to be a simple consequence of the formalism.
\end{abstract}
\vskip3cm
\centerline{\hfill In memory of our friend Stanislaw Malinowski  }
\newpage
The Calogero model~[1],[2],[3],[4], although introduced more than a
quater of  
century ago, still attracts much attention. It has been show to be
related to many  
branches of theoretical physics like the theory of quantum Hall
effect~[5], fractional 
statistics~[6], two--dimensional gravity~[7], two--dimensional QCD~[8]
and others.  
Many advanced techniques has been applied in order to shed light on
the structure 
of the model: inverse scattering method~[3],[9], r--matrix
methods~[10], W--algebra techniques~[11] etc.
Many aspects of Calogero model can be understood by fairly elementary
methods. For instance Barucchi 
and Regge~[12] and Wojciechowski~[13] have shown that the $sL(2$,{\Bbb
  R}) algebra plays an important 
role in the structure of Calagero model without harmonic external
potential. In particular, the  
superintegrability of the model~[14] can be easily shown using
elementary group theory. 
In the present note we show how the Calogero model with harmonic term
can be dealt with in a similar 
way using $sL(2$,{\Bbb R}) algebra. We construct explicitly functions
defined 
over phase space with a very simple (exponential) time dependence; in
particular, 
the integrals of motion are given and their involutivness is shown by
refering to the pure 
(i.e. without harmonic term) Calogero model. It follows immediately
from 
our results that the model retains the property of superintegrability
after 
including the harmonic potential. This latter result is known~[3],[15]
but here is  
shown to be a straightforward consequences of $sL(2$,{\bb R})
dynamical symmetry. 
Let us recall the construction of $sL(2$,{\bb R}) algebra for the pure
Calagero model~[13]. 
To any function $f(q,p)$\ we ascribe the operator $\hat{F}$\ acting in
the linear space of functions defined over phase space:
\be
\{f,g\}&=&\hat{F}g
\ee
\ni for any function $g$. Obviously
\be
\hat{F}&=&\sum\limits_i\left(\frac{\partial f}{\partial q_i}
\frac{\partial}{\partial p_i}-\frac{\partial f}{\partial p_i}
\frac{\partial}{\partial q_i}\right).
\ee
Let us define the following three functions
\be
t_+&=&-\frac{1}{2}\left(\sum\limits^N_{i=1}p^2_i+g^2
\sum\limits^N_{i,j=1}{}'\frac{1}{(q_i-q_j)^2}\right)\nn\\
t_0&=&\frac{1}{2}\sum\limits^N_{i=1}q_ip_i\\
t_-&=&\frac{1}{2}\sum\limits^N_{i=1}q^2_i\nn
\ee
They obey the $sL(2$,{\bb R}) algebra rules (with respect to the standard
Poisson brackets)
\be
\{t_0,t_\pm\}&=&\pm t_\pm\\
\{t_+,t_-\}&=&2t_0\nn
\ee
\ni or, respectively
\begin{eqnarray}
[\hat{T}_0,\hat{T}_\pm]&=&\pm \hat{T}_\pm 
\ee
\be
[\hat{T}_+,\hat{T}_0]&=&2\; \hat{T}_0,\nn
\end{eqnarray}
\ni where, according to eq.~(1)
\be
\hat{T}_+&=&\sum\limits^N_{i=1}p_i\frac{\partial}{\partial q_i}+
\frac{g^2}{2}\sum\limits^N_{i,j=1}
\frac{1}{(q_i-q_j)^3}\left(\frac{\partial}{\partial p_j}-
\frac{\partial}{\partial p_i}\right)\nn  \\
\hat{T}_0&=&\frac{1}{2}\sum\limits^N_{i=1}\l(p_i
\frac{\partial}{\partial p_i}-q_i\frac{\partial}{\partial q_i}\r)\\
\hat{T}_-&=&\sum\limits^N_{i=1}q_i\frac{\partial}{\partial p_i}\nn
\ee
The equations of motion for the Calogero model can be written as
\be
\frac{df}{dt}=&\{f,-t_+\}&=\hat{T}_+f
\ee
Therefore, the integrals of motion are highest--weight vectors; they
can 
be chosen to be (half--) integer eigenvectors of $\hat{T}_0$\ thus
providing 
a finite--di\-men\-sio\-nal irreducible representations of $sL(2$,{\bb
  R}). Moreover, 
elementary group theory allows us to give an immediate proof of
superintegrability 
of rational Calogero model (shown, in somewhat different way,
in~[14]). To this 
end let us note that it is sufficient to find N independent quantities
evolving 
linearly in time (due to the noncompactness of the system no condition
for ratios of  
frequencies~[16] are necessary). But this is rather trival: for if
$f$\ is an integral 
of motion then $\hat{T}_-f$\ depends linearly on time provided $f$\ is
an eigenvector of $\hat{T}_0$.
The $sL(2$,{\bb R}) algebra can be slightly extended. Let us  add two
further functions and corresponding operators
\be
s_+=\sum\limits^N_{i=1}p_i&,&\hat{S}_+=
-\sum\limits^N_{i=1}\frac{\partial}{\partial q_i}\\
s_-=\sum\limits^N_{i=1}q_i&,&\hat{S}_-=
\sum\limits^N_{i=1}\frac{\partial}{\partial p_i}\nn
\ee
The elements $t_0,t_\pm,s_\pm$\ obey the following algebra
\be
\{t_0,s_\pm\}&=&\pm\frac{1}{2}s_\pm\nn\\
\{t_\pm,s_\pm\}&=&0\nn\\
\{t_\pm,s_\mp\}&=&s_\pm\\
\{s_-,s_+\}&=&N\nn
\ee
However, for the operators $\hat{S}_\pm$\ the last formula is to be
replaced by 
\be
[\hat{S}_-,\hat{S}_+]&=&0
\ee
The algebra of operators $\hat{T}_0,\hat{T}_\pm,\hat{S}_\pm$\ is
therefore a  
semidirect product of $sL(2$,{\bb R}) with two--dimensional abelian
algebra 
spanned by $sL(2$,{\bb R}) dublet $\hat{S}_\pm$. The Poisson
algebra~(9) provides 
a central extension of the latter, the parameter of extension being
the number of particles N.
The representations of the algebra under consideration can be easily
obtained. 
We describe the simplest one containing all independent integrals of
motion. 
Let $f_{00}$ be  the highest--weight vector such that
$\hat{T}_of_{00}=\frac{N}{2}f_{00}, 
\hat{S}_+f_{00}=0$. One can take, for example, the
translation--invariant integral 
of motion for the Calogero model given in~[17]~[18]
\be
f_{00}&=&e^{\frac{-g}{2}\sum\limits^N_{i,j=1}{}'\frac{1}{(q_i-q_j)^2}
\frac{\partial^2}{\partial p_i\partial p_j}}\prod^N_{k=1}p_k
\ee 
Define the vectors
\be
f_{mn}\equiv\hat{T}_-^m\hat{S}_-^nf_{00},&
0\leq m\leq N-n,&0\leq n\leq N-1;
\ee
they span a subspace carrying an irreducible representation of our
algebra. It reads  
\be
\hat{T}_-f_{mn}&=&f_{m+1\,n}\nn\\     
\hat{S}_-f_{mn}&=&f_{m\,n+1}\nn\\
\hat{T}_0f_{mn}&=&\l(\frac{N}{2}-m-\frac{n}{2}\r)f_{mn}\\
\hat{S}_+f_{mn}&=&-mf{_m-1\,n+1}\nn\\
\hat{T}_+f_{mn}&=&m\l(N-m-n+1\r)f_{m-1\,n}\nn
\ee
In particular, it follows from the above formulae that
$f_{0n},n=0,\ldots,N-1$\  
are  translation--invariant integrals of motion for Calogero
model. They are obviously 
linearly independent; however, their functional independence can be
checked only by direct inspection.\\
Also
\be
\hat{T}_+f_{1n}&=&(N-n)f_{0n}
\ee
implies that 
\be
(N-k)f_{0k}f_{1n}-(N-n)f_{0n}f_{1k}&&
\ee
\ni are again integrals of motion. Obviously, only at most $N-1$\ of
them can be 
independent and, also by direct inspection, we verify that this is
actually the case.
Let us pass to our main theme --- the Calogero model in external
harmonic oscillator potential. The hamiltonian of the model reads
\be
H&=&\frac{1}{2}
\l(\sum\limits^N_{i=1}p^2_i+g\sum\limits^N_{i,j=1}%
{}'\frac{1}{(q_i-q_j)^2}+\omega^2\sum\limits^N_{i=1}q^2_i\r)
\ee
Under the redefinition $t_0\to t_0, t_\pm\to\omega^{\mp 1}t_\pm$\  
the $sL(2$,{\bb R}) algebra remains unchanged. Our hamiltonian can be
written as
\be
H=&\omega (\hat{T}_--\hat{T}_+)&=-2i\omega\hat{T}_2
\ee
The $\omega \ne 0$\ case differs qualitatively from the $\omega=0$\  one.
On the algebraic level this is reflected in the difference in spectral
properties of $\hat{T}_+$\ and $\hat{T}_2$. Group theory allows us to
find easily the functions having simple time behavior. Let $e_m^s ,
m=-s,\ldots,s$\  
be a basis of spin $s$\ representation of $sL(2,$\rr); the
normalization convention adopted is such that 
\be
\hat{T}_+e_m^s&=&(s-m)(s+m+1)e^s_{m+1}\nn\\
\hat{T}_-e^s_m&=&e^s_{m-1}.\nn
\ee
Then the equations
\be
(\hat{T}_--\hat{T}_+)\phi^s_k&=&-2ik\phi^s_k, k=-s,\ldots,s\nn\\
\hat{T}^2\phi^s_k&=&s(s+1)\phi^s_k
\ee
imply
\be
\phi^s_k=\sum\limits^s_{m=-s}\frac{(s+m)!}{(2s)!}c_{s+m}(s,k)e^s_m
\ee
where the coefficients $c_n(s,k)$\ are defined by 
\be
(x+i)^{s+k}(x-i)^{s-k}&=&\sum\limits^{2s}_{n=0}c_n(s,k)x^n
\ee
Now, it is easy to solve the Hamilton equations for $\phi^s_k$:
\be
\frac{d\phi^s_k}{dt}=&\left\{\phi^s_k, 
\omega(t_--t_+)\right\}&=-\omega(\hat{T}_--\hat{T}_+)\phi^s_k
\ee
give
\be
\phi^s_k(t)&=&e^{2ik\omega t}\phi^s_k(0)
\ee
Let us take as $e^s_s$\ the integral~(11), $N=2s$. Writing
equation~(19)  
in the form
\be
\phi^s_k&=&\sum\limits^{2s}_{n=0}
\frac{(2s-n)!}{(2s)!}
c_{2s-n}(s,k)\l(\omega\sum\limits^N_{i=1}q_i
\frac{\partial}{\partial p_i}\r)^ne^s_s
\ee
and using eqs.~(11) and~(20) one finds\\ 
\newpage
\addtocounter{equation}{1}
$\phi^s_k=\frac{1}{(2s)!}
e^{-\frac{g}{2}\sum\limits^N_{i,j=1} {}'\frac{1}{(q_i-q_j)^2}
\frac{\partial^2}%
{\partial p_i\partial p_j}}F(q,p;\omega)$\hfill (24a)\\
\vs0.3cm
\ni$F(q,p;\omega)=
\sum\limits^{}_{\stackrel{\delta\subset\{1,\ldots,2s\}}{|\delta|=s+k}}
\prod\limits^{}_{i\in\delta}(p_i+i\omega q_i)
\prod\limits^{}_{i\not\in\delta}(p_i-i\omega q_i)$\hfill (24b)\\
The function $\phi^{s'}_k, s'<s$, can be obtained by taking, for
example, 
$e^{s'}_{s'}=f_{02(s-s')}$. Let us note that $f_{02(s-s')}$\ has the
following 
form: one chooses a subsystem consisting of $N'=2s'$\ particles and
construct 
the relevant integral~(11); $f_{02(s-s')}$\ is the sum of such
expressions over 
all choices of subsystems of $N'$\ particles. It is readily seen from
our construction that $\phi^{s'}_{k'}, s\le s'$, have the same
structure. 
In this way we obtained an explicit representation of functions which
have a simple time behaviour under the hamiltonian flow generated by
the hamiltonian of Calogero model in external harmonic potential.
In order to show complete integrability of Calogero model with
harmonic potential 
it is more convenient to start with $e^s_s$ given by another
well--known formula 
for pure $(\omega=0$) Calogero model integrals of motion
\be
e^s_s&=&Tr(L^{2s})
\ee
where $L$\ is the relevant Lax matrix. Taking $k=0, s=1,\ldots,N$\ and
inserting  
$e^s_s$\ as given by eq.~(25) into eq.~(23) 
we obtain $N$\ integrals of motion.
\vs0.3cm
\be
\phi^s_0&=&\sum\limits^s_{m=0}
\frac{(2s-2m)!}{(2s)!}{s \atopwithdelims () m}
\omega^{2m}\l(\sum\limits^N_{i=1}q_i
\frac{\partial}{\partial p_i}\r)^{2m}Tr(L^{2s})
\ee
\vs0.3cm
In order to prove that they are in involution, let us first note that
\vs0.3cm
\be
e^s_0&\simeq&\l(\sum\limits^N_{i=1}q_i
\frac{\partial}{\partial p_i}\r)^sTr(L^{2s}), s=1,\ldots,N
\ee
\vs0.3cm
are in involution~[19]. Moreover, we have the following identity
\vs0.3cm
\be
e^{\frac{i\pi}{4}(\hat{T}_++\hat{T}_-)}
\hat{T}_0e^{-\frac{i\pi}{4}(\hat{T}_++\hat{T}_-)}
&=&\frac{i}{2}(\hat{T}_--\hat{T}_+)
\ee
\vs0.3cm
Therefore
\vs0.3cm
\be
\phi^s_0&\simeq&e^{\frac{i\pi}{4}(\hat{T}_++\hat{T}_-)}e^s_0
\ee
\vs0.3cm
However, $\omega(\hat{T}_++\hat{T}_-)$\ is the operator representing the
function
\vs0.3cm
\be
-h&=&-\l(\frac{1}{2}\sum\limits^N_{i=1}p_i^2+
\frac{g}{2}\sum\limits^N_{i,j=1}{}'\frac{1}{(q_i-q_j)^2}-
\frac{\omega^2}{2}\sum\limits^N_{i=1}q^2_i\r)
\ee
which is the hamiltonian for Calogero model in inverse harmonic
potential. 
Therefore, it follows from eq.~(29) 
that $\phi^s_0$\ is obtained from $e^s_0$\ evolving during the time 
$\frac{i\pi}{4\omega}$\ (which is purely imaginary but this is
irrelevant 
in what follows) according to the hamiltonian flow determined by $h$.
The time evolution is a canonical transformation which proves that
$\phi^s_0$\ are also in involution.
In order to show that the integrals~(26) 
are independent for $s=1,\ldots,N$\ it is sufficient to note that they
contain only even powers of momenta and have the form
\be
\phi^s_0&=&\sum\limits^N_{i=1}p_i^{2s}+
{\rm terms\;\; of\;\; lower\;\; degree\;\; in\;\; p's}
\ee
It is also easy to check that for $g=0$\ they reduce to the following
ones 
\be
\phi^s_0(g=0)&=&\sum\limits^N_{i=1}(p_i^2+\omega^2q^2_i)^s , s=1,\ldots,N
\ee
The superintegrability of the model can be also shown by our
method. This is fairly obvious -- we have constructed a huge set of
functions depending harmonically on time, the ratios of frequencies
being rational numbers. 
To conclude we have shown that (super-) integrability of the Calogero
model with harmonic external potential can be easily derived from the
properties of pure Calogero model by using elementary group
theoretical techniques. 
 
\end{document}